\documentclass[aps,onecolumn,showpacs,nofootinbib,showkeys]{revtex4-2}
\usepackage[margin=3.15cm]{geometry}

\RequirePackage[T1]{fontenc}

\usepackage{epsfig,graphicx,amsmath,amssymb,bm}
\RequirePackage{mathptmx}      
\usepackage{mathrsfs}
\usepackage{amssymb}
\RequirePackage{color}
\usepackage{changes}
\usepackage{slashed}
\usepackage{multirow}
\usepackage{scalerel}
\usepackage{tikz-feynman}
\usepackage{textcomp}
\usepackage{subcaption}
\usepackage{float}
\RequirePackage{hyperref}
\hypersetup{
    linktocpage,
    colorlinks,
    citecolor=blue,
    filecolor=black,
    linkcolor=blue,
    urlcolor=blue,
}
\usepackage{nccmath}

\begin{document}

\title{A theoretical analysis of the semileptonic decays\\
$\eta^{(\prime)}\to\pi^0l^+l^-$ and $\eta^\prime\to\eta l^+l^-$}


\author{Rafel Escribano$^{1,2}$}\email{rescriba@ifae.es}
\author{Emilio Royo$^{1,2}$}\email{eroyo@ifae.es}

\affiliation{$^1$Grup de F\'{i}sica Te\`orica, Departament de F\'{i}sica, Universitat Aut\`onoma de Barcelona, E-08193 Bellaterra (Barcelona), Spain\\
$^2$Institut de F\'{i}sica d'Altes Energies (IFAE), The Barcelona Institute of Science and Technology, Campus UAB, E-08193 Bellaterra (Barcelona), Spain}   



\begin{abstract}
A complete theoretical analysis of the $C$-conserving semileptonic decays 
$\eta^{(\prime)}\to\pi^0l^+l^-$ and $\eta^\prime\to\eta l^+l^-$ ($l=e$ or $\mu$)
is carried out within the framework of the Vector Meson Dominance (VMD) model.
An existing phenomenological model is used to parametrise the VMD 
coupling constants and the associated numerical values are 
obtained from an optimisation fit to $V\to P\gamma$
and $P\to V\gamma$ radiative decays
($V=\rho^0$, $\omega$, $\phi$ and $P=\pi^0$, $\eta$, $\eta^{\prime}$).
The decay widths and dilepton energy spectra for the two $\eta\to\pi^0l^+l^-$ processes
obtained using this approach are compared and found to be in good agreement with
other results available in the published literature. 
Theoretical predictions for the four $\eta^{\prime}\to\pi^0l^+l^-$ and
$\eta^\prime\to\eta l^+l^-$ decay widths and dilepton energy spectra
are calculated and presented for the first time in this work.


\end{abstract}


\maketitle


\section{\label{Intro}Introduction}

The electromagnetic and strong interactions conserve parity ($P$) and charge conjugation ($C$) 
within the well-established and well-tested Standard Model of particle physics (SM).
In this framework, the $\eta$ and $\eta^{\prime}$ pseudoscalar mesons are specially suited for 
the study of rare decay processes, for instance, in search of $C$, $P$ and $CP$ violations,
as these mesons are $C$ and $P$ eigenstates of the electromagnetic and strong interactions 
\cite{Gan:2020aco}. 

Specifically, the semileptonic decays $\eta^{(\prime)}\to\pi^0l^+l^-$ and 
$\eta^\prime\to\eta l^+l^-$ ($l=e$ or $\mu$) are of special interest given that
they can be used as fine probes to assess if new physics beyond the Standard Model (BSM) is at play.
This is because any contribution from BSM physics ought to be relatively small 
and the above decay processes only get a contribution from the SM through the $C$-conserving 
exchange of two photons that is highly suppressed,
as there is no contribution at tree-level but only corrections at one-loop and higher orders. 
This small SM contribution would presumably be of the same order of magnitude as that of 
physics BSM, which, in turn, means that the $\eta$-$\eta^{\prime}$ phenomenology might play
an interesting role and be an excellent arena for stress testing SM 
predictions \cite{Gan:2020aco,Fang:2017qgz}.
As an example, the $\eta^{(\prime)}\to\pi^0l^+l^-$ and $\eta^\prime\to\eta l^+l^-$ decays
could be mediated by a single intermediate virtual photon,
but this would entail that the electromagnetic interactions violate $C$-invariance 
(e.g.~\cite{Bernstein:1965hj,Lee:1965zz}) and, therefore,
would represent a departure from the SM.

Early theoretical studies of semileptonic decays of pseudoscalar 
mesons date back to the late 1960s.
A very significant contribution was made by Cheng in Ref.~\cite{Cheng:1967zza}
where he analysed the $\eta\to\pi^0e^+e^-$ decay mediated by a $C$-conserving,
two-photon intermediate state within the Vector Meson Dominance (VMD) framework.
By setting the electron mass to $m_e=0$ and neglecting 
in the numerator of the amplitude
terms that were second or higher order in the electron or positron $4$-momenta,
he found theoretical estimations for the decay width 
$\Gamma(\scaleto{\eta\to\pi^0}{10pt}e^+e^-)=1.3\times10^{-5}$ eV,
the relative branching ratio
$\Gamma(\scaleto{\eta\to\pi^0}{10pt}e^+e^-)/\Gamma(\scaleto{\eta\to\pi^0\gamma\gamma}{10pt})
\approx10^{-5}$, as well as the associated decay energy spectrum. 
This was an enormous endeavour given the very limited access 
to computer algebra systems at the time.
For this reason, a number of strong assumptions had to be made, 
as pointed out above, which may have had an undesired effect 
on the accuracy of Cheng's estimates.
A different approach was followed by Smith~\cite{Smith:1968erc} 
also in the late 1960s,
whereby an $S$-wave $\eta\pi^0\gamma\gamma$ coupling and unitary bounds\footnote{As
it is well known, the Cutkovsky rules~\cite{Cutkosky:1960sp} allow one to calculate the
imaginary part of a transition amplitude by putting the intermediate virtual particles on-shell.}
were used for the calculation of the $C$-conserving modes associated to both
$\eta\to\pi^0l^+l^-$ decay processes.
By neglecting $p$-wave contributing terms to simplify the calculations and noting that
the unknown $\eta\pi^0\gamma\gamma$ coupling constant cancels out when calculating
relative branching ratios, Smith was able to find
$\Gamma(\scaleto{\eta\to\pi^0}{10pt}e^+e^-)/
\Gamma(\scaleto{\eta\to\pi^0\gamma\gamma}{10pt})=3.6\times10^{-8}$ and
$\Gamma(\scaleto{\eta\to\pi^0\mu^+\mu^-}{10pt})/
\Gamma(\scaleto{\eta\to\pi^0\gamma\gamma}{10pt})=6.0\times10^{-5}$,
after estimating the real part of the matrix element from a single dispersion relation
and employing a cut-off $\Lambda = 2m_{\eta}$.
The calculation of the latter ratio was possible due to the fact that Smith
did not approximate the lepton mass to zero.

Ng et al.~\cite{Ng:1992yg} also found in the early 1990s 
lower limits for the decay widths
of the two $\eta\to\pi^0l^+l^-$ processes by making use of 
unitary bounds and the decay chain
$\eta\to\pi^0\gamma\gamma\to\pi^0l^+l^-$.
The transition form factors associated to the $\eta\to\pi^0\gamma\gamma$ decay,
which are required to perform the above calculation,
were obtained using the VMD model supplemented by the 
exchange of an $a_0$ scalar meson.
The lower bounds that they found are
$\Gamma(\scaleto{\eta\to\pi^0}{10pt}e^+e^-)\vert_{\scaleto{\textrm{VMD}}{4pt}}=
1.1_{\scaleto{-0.5}{5pt}}^{\scaleto{+0.6}{5pt}}\ \mu$eV and
$\Gamma(\scaleto{\eta\to\pi^0\mu^+\mu^-}{10pt})\vert_{\scaleto{\textrm{VMD}}{4pt}}=
0.5_{\scaleto{-0.2}{5pt}}^{\scaleto{+0.3}{5pt}}\ \mu$eV, making use of VMD only.
By adding the $a_0$ exchange\footnote{The
$a_0\eta\pi^0$ and $a_0\gamma\gamma$ couplings needed 
to perform this calculation
were roughly estimated and the authors acknowledged to be poorly known.
As well as this, their signs were not unambiguously fixed.}
to the latter process, they obtained
$\Gamma(\scaleto{\eta\to\pi^0\mu^+\mu^-}{10pt})\vert_{\scaleto{\textrm{constr}}{4pt}}=
0.9_{\scaleto{-0.5}{5pt}}^{\scaleto{+0.6}{5pt}}\ \mu$eV and
$\Gamma(\scaleto{\eta\to\pi^0\mu^+\mu^-}{10pt})\vert_{\scaleto{\textrm{destr}}{5pt}}=
0.3_{\scaleto{-0.2}{5pt}}^{\scaleto{+0.4}{5pt}}\ \mu$eV
for a constructive and destructive interference, respectively.
The real parts of the amplitudes were estimated 
by means of a cut-off dispersive relation
and the authors argued that the expected dispersive 
contribution should be no larger than
30\% of the absorptive one.
A few months later, Ng and Peters provided in 
Ref.~\cite{Ng:1993sc} new estimations
for the unitary bounds of the $\eta\to\pi^0l^+l^-$ decay widths.
This new contribution was two-fold; on one hand,
they calculated the $\eta\to\pi^0\gamma\gamma$ decay width within a
constituent quark model framework; on the other hand,
they recalculated the VMD transition form factors from Ref.~\cite{Ng:1992yg}
by performing a Taylor expansion and keeping terms linear in
$M_{\eta}^2/M_V^2$, $x_1$ and $x_2$ ($x_i\equiv P_{\eta}\cdot q_{\gamma_i}/M_{\eta}^2$),
which had been neglected in their previous work.
Their new findings were:~(i)
$\Gamma(\scaleto{\eta\to\pi^0}{10pt}e^+e^-)\vert_{\scaleto{\textrm{box}}{4pt}}
\geq1.2\pm0.2\ \mu$eV and
$\Gamma(\scaleto{\eta\to\pi^0\mu^+\mu^-}{10pt})\vert_{\scaleto{\textrm{box}}{4pt}}
\geq4.3\pm0.7\ \mu$eV for a constituent quark mass $m=330$ MeV$/c^2$; and (ii)
$\Gamma(\scaleto{\eta\to\pi^0}{10pt}e^+e^-)\vert_{\scaleto{\textrm{VMD}}{4pt}}
\geq3.5\pm0.8\ \mu$eV and
$\Gamma(\scaleto{\eta\to\pi^0\mu^+\mu^-}{10pt})\vert_{\scaleto{\textrm{VMD}}{4pt}}
\geq2.4\pm0.8\ \mu$eV.
It is important to highlight that their estimations using the quark-box mechanism were
strongly dependent on the specific constituent quark mass selected,
especially for the electron mode.
\begin{figure*}[t]
 \centering
  \begin{subfigure}{0.5\textwidth}
   \centering
    \begin{tikzpicture}
     \begin{feynman}
      \vertex (a) {\(\eta^{(\prime)}\)};
      \vertex [dot, right=2cm of a] (b) {};
      \vertex [dot, above right=2cm of b] (c) {};
      \vertex [dot, below right=2cm of b] (d) {};
      \vertex [right=2.5cm of c] (g) {\(l^{+}\)};
      \vertex [dot, above right=2cm of d] (e) {};
      \vertex [right=2cm of e] (h) {\(l^{-}\)};
      \vertex [right=2.5cm of d] (f) {\(\pi^{0} (\eta)\)};
      \diagram* {
        (a) -- [scalar, momentum={[arrow shorten=0.08mm, arrow distance=1.5mm, 
        inner sep=1.5pt]\(P\)}] (b),
        (b) -- [photon, near start, inner sep=1pt, edge label'=\(\gamma^{*}_1\), 
        momentum={[arrow shorten=0.1mm, arrow distance=2mm, inner sep=1pt]\(k\)}] (c),
        (d) -- [edge label'=\(V\), inner sep=1pt, rmomentum={[arrow shorten=0.08mm, 
        arrow distance=1.5mm, inner sep=1pt]\(P-k\)}] (b),
        (d) -- [photon, inner sep=1pt, edge label'=\(\gamma^{*}_2\), 
        momentum={[arrow shorten=0.1mm, arrow distance=2mm, inner sep=1pt]\(q-k\)}] (e),
        (c) -- [anti fermion, momentum={[arrow shorten=0.09mm, arrow distance=2mm, 
        inner sep=1.5pt]\(p_+\)}] (g),
        (e) -- [anti fermion, rmomentum={[arrow shorten=0.08mm, arrow distance=2mm, 
        inner sep=0pt]\(k-p_+\)}] (c),
        (e) -- [fermion, momentum={[arrow shorten=0.08mm, arrow distance=2mm, 
        inner sep=1.5pt]\(p_-\)}] (h),
        (f) -- [scalar, rmomentum={[arrow shorten=0.09mm, arrow distance=1.5mm, 
        inner sep=2.5pt]\(p_0\)}] (d),
      };
     \end{feynman}
    \end{tikzpicture}
   \caption{$t$-channel Feynman diagram}
  \end{subfigure}%
 \centering
 \begin{subfigure}{0.5\textwidth}
  \centering
   \begin{tikzpicture}
    \begin{feynman}
      \vertex (a) {\(\eta^{(\prime)}\)};
      \vertex [dot, right=2cm of a] (b) {};
      \vertex [dot, above right=2cm of b] (c) {};
      \vertex [dot, below right=2cm of b] (d) {};
      \vertex [below right=1.7cm and 3.3cm of c] (g) {\(l^{-}\)};
      \vertex [dot, above right=2cm of d] (e) {};
      \vertex [above right=2.2cm of e] (h) {\(l^{+}\)};
      \vertex [right=2.5cm of d] (f) {\(\pi^{0} (\eta)\)};
      \diagram* {
        (a) -- [scalar, momentum={[arrow shorten=0.08mm, arrow distance=1.5mm, 
        inner sep=1.5pt]\(P\)}] (b),
        (b) -- [photon, near start, inner sep=1pt, edge label'=\(\gamma^{*}_1\), 
        momentum={[arrow shorten=0.1mm, arrow distance=2mm, inner sep=1pt]\(k\)}] (c),
        (d) -- [edge label'=\(V\), inner sep=1pt, rmomentum={[arrow shorten=0.08mm, 
        arrow distance=1.5mm, inner sep=1pt]\(P-k\)}] (b),
        (d) -- [photon, inner sep=1pt, edge label'=\(\gamma^{*}_2\), 
        momentum={[arrow shorten=0.1mm, arrow distance=2mm, inner sep=1pt]\(q-k\)}] (e),
        (c) -- [fermion, inner sep=0pt, bend left, momentum={[arrow shorten=0.12mm, 
        arrow distance=2mm, inner sep=1.5pt]\(p_-\)}] (g),
        (e) -- [fermion, momentum={[arrow shorten=0.08mm, arrow distance=2mm, 
        inner sep=0pt]\(p_{-}-k\)}] (c),
        (h) -- [fermion, inner sep=1pt, bend left, rmomentum={[arrow shorten=0.09mm, 
        arrow distance=2mm, inner sep=1.5pt]\(p_+\)}] (e),
        (f) -- [scalar, rmomentum={[arrow shorten=0.09mm, arrow distance=1.5mm, 
        inner sep=2.5pt]\(p_0\)}] (d),
      };
     \end{feynman}
    \end{tikzpicture}
   \caption{$u$-channel Feynman diagram}
  \end{subfigure}%
 \caption{Feynman diagrams contributing to the $C$-conserving semileptonic decays 
 $\eta^{(\prime)}\to\pi^0l^+l^-$ and $\eta^{\prime}\to\eta l^+l^-$ ($l=e$ or $\mu$).
 Note that $q=p_{+}+p_{-}$ and 
 $V=\scaleto{\rho^0}{9.5pt},\scaleto{\omega}{5pt},\scaleto{\phi}{9pt}$.}
 \label{fig1}
\end{figure*}
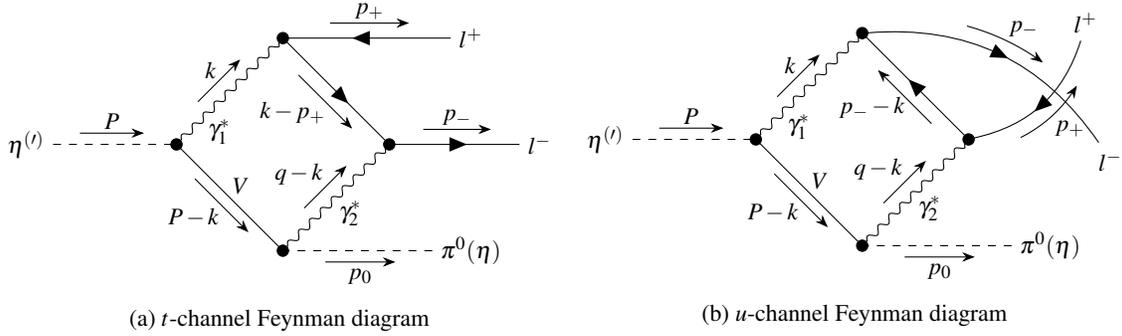%

On the experimental front,
new upper limits have recently been established by the 
WASA-at-COSY collaboration for the 
$\eta\to\pi^0e^+e^-$ decay width~\cite{Adlarson:2018imw}.
This is a useful contribution, 
as the previous available empirical measurements date back to the 1970s which provided
an upper limit for the relative branching ratio of the above process that was
many orders of magnitude larger than the corresponding theoretical 
estimations at the time.
In particular, Adlarson et al.~\cite{Adlarson:2018imw} found from the analysis
of a total of $3\times 10^7$ events of the reaction 
$\textrm{pd}\to \!^3\!\textrm{He}+\eta$,
with a recorded excess energy of $Q=59.8$ MeV,
that the results are consistent with no $C$-violating 
single-photon intermediate state event 
being recorded.
Based on their analysis, the new upper limits 
$\Gamma(\scaleto{\eta\to\pi^0}{10pt}e^+e^-)/\Gamma(\scaleto{\eta\to\pi^+\pi^-\pi^0}{10pt})
<3.28\times10^{-5}$ and
$\Gamma(\scaleto{\eta\to\pi^0}{10pt}e^+e^-)/\Gamma(\scaleto{\eta\to\textrm{all}}{8pt})
<7.5\times10^{-6}$ (CL $=90\%$)
have been established for the $C$-violating $\eta\to\pi^0\gamma^*\to\pi^0e^+e^-$ decay.
In addition, the WASA-at-COSY Collaboration is currently analysing additional data from the 
$\textrm{pp}\to\textrm{pp}\eta$ reaction collected over three periods in 2008, 2010 and 2012 
which should put more stringent upper limits on the $\eta\to\pi^0e^+e^-$ branching ratio.
The experimental state of play is expected to be further improved in the near future with the 
advent of new experiments such as the REDTOP,
which will focus on rare decays of the $\eta$ and $\eta^{\prime}$ mesons,
providing increased sensitivity in the search for violations of SM symmetries by
several orders of magnitude beyond the current experimental state of the art 
\cite{Gatto:2019dhj}.

The present work is structured as follows:
In section~\ref{Calc}, we present the detailed calculations for the decay widths associated to 
the six $\eta^{(\prime)}\to\pi^0l^+l^-$ and $\eta^{\prime}\to\eta l^+l^-$ processes.
In section~\ref{Results}, numerical results from theory for the decay widths and
the corresponding dilepton energy spectra are presented and discussed for the six reactions.
Some final remarks and conclusions are given in section~\ref{Conclusions}.


\section{\label{Calc}Calculations of
$\pmb{\eta^{(\prime)}\to\pi^0 l^+ l^-}$ and $\pmb{\eta^{\prime}\to\eta l^+ l^-}$}

The calculations in this work assume that the
$\eta^{(\prime)}\to\pi^0l^+l^-$ and $\eta^{\prime}\to\eta l^+l^-$ decays processes
are dominated by the exchange of vector resonances\footnote{It is
worth highlighting that contributions from the exchange of scalar 
resonances can be safely 
discarded as they ought to be negligible for the first four 
$\eta^{(\prime)}\to\pi^0l^+l^-$ decays
and relatively small for the last two $\eta^{\prime}\to\eta l^+l^-$ processes.
The interested reader is referred to the in-depth analysis carried out in
Ref.~\cite{Escribano:2018cwg} where scalar exchanges were introduced under the framework of 
the Linear Sigma Model for the $\eta^{(\prime)}\to\pi^0\gamma\gamma$ 
and $\eta^{\prime}\to\eta \gamma\gamma$ decays.} \cite{Cheng:1967zza}; that is, they proceed through the
$C$-conserving virtual transition $\eta^{(\prime)}\to V\gamma^*$
(with $V = \rho^0$, $\omega$ or $\phi$), followed by $V\to\pi^0\gamma^*$ (or $V\to\eta\gamma^*$) 
and $2\gamma^{*}\to l^+l^-$ (see Fig.~\ref{fig1} for details)\footnote{Note that
any $C$-violating contributions to these processes,
such as e.g.~the single-photon exchange channel, would be associated to BSM physics.
In this work, though, the focus is on the SM contribution from the $C$-conserving
two-photon exchange channel.}.

In order to perform the calculations,
one first needs to select an effective vertex that contains the appropriate interacting terms.
The $VP\gamma$ interaction amplitude consistent with Lorentz, $P$, $C$ and
electromagnetic gauge invariance can be written as \cite{Prades:1993ys}
\begin{equation}
\begin{aligned}
\mathscr{M}(V\to P\gamma) = g_{VP\gamma}\,\epsilon_{\mu \nu \alpha \beta}
\epsilon^\mu_{(V)}p_V^\nu\epsilon^{\ast\alpha}_{(\gamma)}q^\beta\hat F_{VP\gamma}(q^2)\ ,
\label{eqeffvertex}
\end{aligned}
\end{equation}
where $g_{VP\gamma}$ is the coupling constant for the $VP\gamma$ transition involving on-shell photons, $\epsilon_{\mu \nu \alpha \beta}$ is the totally antisymmetric Levi-Civita tensor, $\epsilon_{(V)}$ and $p_V$ are the polarisation and 4-momentum vectors of the initial $V$, $\epsilon_{(\gamma)}^\ast$ and $q$ are the corresponding ones for the final $\gamma$, 
and $\hat F_{VP\gamma}(q^2)\equiv F_{VP\gamma}(q^2)/F_{VP\gamma}(0)$ is a normalised form factor to account for off-shell photons mediating the transition\footnote{For simplicity of the calculation,
we neglect the $q^2$ dependence of the transition form factor
in Eq.~(\ref{eqeffvertex}).
This is not fully rigorous but, we understand, it is a tolerable approximation
given that these form factors are usually determined from on-shell photon processes.}.
In addition to this, the usual QED vertex is used to describe the subsequent
$2\gamma^{*}\to l^+l^-$ transition. Accordingly, there are six diagrams (two per vector meson) contributing to each one of the six semileptonic decay processes and the corresponding Feynman diagrams are shown in Fig.~\ref{fig1}.

The invariant decay amplitude in momentum space can, therefore, be written as follows

\begin{widetext}
\begin{equation}
\begin{aligned}
\mathcal{M} = ie^2 \sum_{V=\rho^0,\omega,\phi} g_{V\eta^{(\prime)}\gamma}\ & 
g_{V\pi^0(\eta)\gamma}\int 
\frac{d^4k}{(2\pi)^4}\frac{1}{k^2+i\scaleto{\epsilon}{5pt}}\frac{1}{(k-q)^2
+i\scaleto{\epsilon}{5pt}}\epsilon_{\mu \nu \alpha \beta}
\Bigg[\frac{k^{\mu}(P-k)^{\alpha}(k-q)^{\rho}(P-k)^{\delta}}{(P-k)^2-m_V^2
+i\scaleto{\epsilon}{5pt}}\Bigg]\epsilon_{\rho \sigma \delta}{}^{\beta} \\
& \overline{u}(p_{-})\Bigg[\gamma^{\sigma}
\frac{\slashed{k}-\slashed{p}_{+}+m_l}{(k-p_{+})^2-m_l^2+i\scaleto{\epsilon}{5pt}}\gamma^{\nu}+
\gamma^{\nu}\frac{\slashed{p}_{-}-\slashed{k}+m_l}{(k-p_{-})^2-m_l^2+i\scaleto{\epsilon}{5pt}}
\gamma^{\sigma}\Bigg]v(p_{+})\ ,
\end{aligned}
\label{eqamp1}
\end{equation}
\end{widetext}

\noindent
where $q=p_{+}+p_{-}$ is the sum of lepton-antilepton pair $4$-momenta,
$e$ is the electron charge, and
$g_{V\eta^{(\prime)}\gamma}$ and $g_{V\pi^0(\eta)\gamma}$
are the corresponding VMD coupling constants in Eq.~(\ref{eqeffvertex}).
Noting that the Levi-Civita tensors are antisymmetric under the substitutions 
$\scaleto{\mu}{7pt}\leftrightarrow\scaleto{\alpha}{5pt}$ and 
$\scaleto{\rho}{7pt}\leftrightarrow\scaleto{\delta}{7pt}$,
whilst the products of loop momenta $k^{\mu}k^{\alpha}$ and $k^{\rho}k^{\delta}$
are symmetric under these substitutions,
one finds that the terms in Eq.~(\ref{eqamp1}) containing these combinations vanish
and that the superficial degree of divergence for the loop integrals of the two diagrams
in Fig.~\ref{fig1} is $-1$.
Accordingly, both diagrams are convergent individually.

The numerator of $\mathcal{M}$ can be simplified using the usual Dirac algebra manipulations
and the equations of motion.
For these calculations, the mass of the leptons are not approximated to zero,
as we are interested in both the electron and muon modes for the six decay processes;
as a result, the task of manipulating and simplifying the algebraic expressions
would be daunting should computer algebra packages not be available.
In the present work, use of the Mathematica package FeynCalc 9.2.0
\cite{Shtabovenko:2016sxi,Mertig:1990an} is made for this purpose. 

Let us now proceed to calculate the loop integral.
As usual, one first introduces the Feynman parametrisation and completes the square
in the new denominators $\Delta_{iV}$ ($i=1,2$ and
$V=\scaleto{\rho^0}{9.5pt},\scaleto{\omega}{5pt},\scaleto{\phi}{9pt}$)
by shifting to a new loop momentum variable $\ell$ \cite{Peskin:1995ev}.
Hence, the denominators become
\begin{equation}
\begin{aligned}
\Delta_{1V} &= 2yz(P\cdot q)+2xy(p_{+}\cdot q)+(y-1)yq^2
+2x z(P\cdot p_{+})+x^2m_{l}^2+z\big[(z-1)m_{\eta^{(\prime)}}^2
+m_V(m_V-i\Gamma_V)\big]\ , \\[6pt]
\Delta_{2V} &\equiv \Delta_{1V} \ \ \mathrm{with} \ \ p_{+}\leftrightarrow p_{-}\ .  
\end{aligned}
\label{eqDeltas12}
\end{equation}

Rewriting the numerators of the Feynman diagrams $1$ and $2$
(i.e.~$t$-channel and $u$-channel diagrams, respectively, in Fig.~\ref{fig1})
in terms of the new momentum variable $\ell$, one finds
\begin{equation}
\begin{aligned}
\mathcal{N}_1 = \big[A_1\ell^2&+B_1\big]\overline{u}(p_{-})\slashed{P}v(p_{+})
+\ m_{l}\big[C_1\ell^2+D_1\big]\overline{u}(p_{-})v(p_{+})\ , \\[6pt]
\mathcal{N}_2 = \big[A_2\ell^2&+B_2\big]\overline{u}(p_{-})\slashed{P}v(p_{+})
+\ m_{l}\big[C_2\ell^2+D_2\big]\overline{u}(p_{-})v(p_{+})\ , 
\end{aligned}
\label{eqnums12}
\end{equation}
where the explicit expressions for the parameters $A_i$, $B_i$, $C_i$ and $D_i$ ($i=1,2$)
are provided in~\ref{appA}.
Finally, we perform a Wick rotation and change to 
four-dimensional spherical coordinates 
\cite{Peskin:1995ev,Schwartz:2013pla} to carry out the momentum integral.
The following expressions for the amplitudes of the Feynman diagrams are found
\begin{equation}
\begin{aligned}
\mathcal{M}_{1_{V}} &=&\!\!\scaleto{\alpha_V}{6pt}
\big[\overline{u}(p_{-})\slashed{P}v(p_{+})\big] + 
\scaleto{\beta_V}{8.5pt}m_l\big[\overline{u}(p_{-})v(p_{+})\big]\ , \\[6pt]
\mathcal{M}_{2_{V}} &=&\!\!\scaleto{\sigma_V}{6pt}
\big[\overline{u}(p_{-})\slashed{P}v(p_{+})\big] + 
\scaleto{\tau_V}{6pt}m_l\big[\overline{u}(p_{-})v(p_{+})\big]\ ,
\end{aligned}
\label{eqamps12}
\end{equation}
where $m_l$ is the corresponding lepton mass, and the parameters 
$\scaleto{\alpha_V}{6pt}$, $\scaleto{\beta_V}{8.5pt}$,
$\scaleto{\sigma_V}{6pt}$ and $\scaleto{\tau_V}{6pt}$ in Eq.~(\ref{eqamps12})
are defined as
\begin{equation}
\begin{aligned}
\scaleto{\alpha_V}{6pt}&=&\!\!\!e^2 \frac{g_{V\eta^{(\prime)}\gamma} 
g_{V\pi^0(\eta)\gamma}}{16\pi^2}\!\int\! dx dy dz 
\Bigg[\frac{2A_1}{\Delta_{1V}\!-\!i\scaleto{\epsilon}{5pt}} - 
\frac{B_1}{(\Delta_{1V}\!-\!i\scaleto{\epsilon}{5pt})^2}\Bigg]\ , \\[2pt]
\scaleto{\beta_V}{8.5pt}&=&\!\!\!e^2 \frac{g_{V\eta^{(\prime)}\gamma} 
g_{V\pi^0(\eta)\gamma}}{16\pi^2}\!\int\! dx dy dz 
\Bigg[\frac{2C_1}{\Delta_{1V}\!-\!i\scaleto{\epsilon}{5pt}} - 
\frac{D_1}{(\Delta_{1V}\!-\!i\scaleto{\epsilon}{5pt})^2}\Bigg]\ , \\[2pt]
\scaleto{\sigma_V}{6pt}&=&\!\!\!e^2 \frac{g_{V\eta^{(\prime)}\gamma} 
g_{V\pi^0(\eta)\gamma}}{16\pi^2}\!\int\! dx dy dz 
\Bigg[\frac{2A_2}{\Delta_{2V}\!-\!i\scaleto{\epsilon}{5pt}} - 
\frac{B_2}{(\Delta_{2V}\!-\!i\scaleto{\epsilon}{5pt})^2}\Bigg]\ , \\[2pt]
\scaleto{\tau_V}{6pt}&=&\!\!\!e^2 \frac{g_{V\eta^{(\prime)}\gamma} 
g_{V\pi^0(\eta)\gamma}}{16\pi^2}\!\int\! dx dy dz 
\Bigg[\frac{2C_2}{\Delta_{2V}\!-\!i\scaleto{\epsilon}{5pt}} - 
\frac{D_2}{(\Delta_{2V}\!-\!i\scaleto{\epsilon}{5pt})^2}\Bigg]\ ,
\end{aligned}
\label{eqFeynpar}
\end{equation}
with $x$, $y$ and $z$ being the Feynman integration parameters.
Therefore, the full amplitude can now be expressed as
\begin{equation}
\begin{aligned}
\mathcal{M} &= \sum_{V=\rho^0,\omega,\phi}\mathcal{M}_{1_{V}}+\mathcal{M}_{2_{V}}
= \Omega\big[\overline{u}(p_{-})\slashed{P}v(p_{+})\big] + 
m_l\Sigma\big[\overline{u}(p_{-})v(p_{+})\big] \ ,
\end{aligned}
\label{eqamp2}
\end{equation}
where $\Omega$ and $\Sigma$ are defined as follows
\begin{equation}
\begin{aligned}
\Omega\ &=& \sum_{V=\rho^0,\omega,\phi}\alpha_V + \sigma_V\ , \\[2pt]
\Sigma\ &=& \sum_{V=\rho^0,\omega,\phi}\beta_V + \tau_V\ ,
\end{aligned}
\label{eqOmSig}
\end{equation}
and the unpolarised squared amplitude is
\begin{equation}
\label{eqsqamp}
\begin{aligned}
\overline{\vert\mathcal{M}\vert^2} &= 4\Big\{2(P\cdot p_{+})(P\cdot 
p_{-})-m_{\eta^{(\prime)}}^2\big[(p_{+}\cdot p_{-})+m_l^2\big]\Big\}
\times \mathrm{Abs}(\Omega)^2+8m_l^2\big[(P\cdot p_{+})-(P\cdot 
p_{-})\big]\mathrm{Re}(\Omega \Sigma^{*})\\
&\quad+ 4m_l^2\big[(p_{+}\cdot p_{-})-m_l^2\big]\mathrm{Abs}(\Sigma)^2\ .
\end{aligned}
\end{equation}
Finally, the differential decay rate for a three-body decay can be written as 
\cite{Zyla:2020zbs}
\begin{equation}
\label{eqdecayrate}
d\Gamma = \frac{1}{(2\pi)^3}\frac{1}{32\ \!m^3_{\eta^{(\prime)}}}
\overline{\vert\mathcal{M}\vert^2}\ dm_{l^+l^-}^2 dm_{l^-\pi^0(\eta)}^2\ ,
\end{equation}
where $m_{ij}^2=(p_i+p_j)^2$.


\section{\label{Results}Theoretical results}

Making use of the theoretical expressions that have been presented in section \ref{Calc}, one can find numerical predictions for the decay widths of the $\eta^{(\prime)}\to\pi^0l^+l^-$ and $\eta^{\prime}\to\eta l^+l^-$ decay processes, as well as their associated dilepton energy spectra.
Both, the integral over the Feynman parameters 
as well as the integral over phase space,
must be carried out numerically, as algebraic expressions cannot be obtained.
In addition, the numerical integrals over the Feynman 
parameters are to be performed using 
adaptive Monte Carlo methods \cite{Press:2007zz};
this is driven by the complexity of the expressions to be integrated and
their multidimensional nature\footnote{It is worth mentioning that comparison between the numerical results for $\Omega$ and $\Sigma$ in Eq.~(8) using the approach presented in this work and Passarino-Veltman reduction techniques implemented in software packages such as, e.g., LoopTools \cite{Hahn:1998yk} 
was carried out for different points of phase space to assess the performance of our method. It was found that our results were in agreement with those from the above package for points far from the edge of phase space, which provides a level of confidence in our approach, but in sharp disagreement for points near the edge of phase space. This is, however, a well-known drawback of the Passarino-Velt\-man reduction and variants due to the appearance of Gram determinants in the denominator, which spoils the numerical stability when they become small or even zero giving rise to spurious singularities (see, e.g., Refs.~\cite{Denner:2005nn,Campbell:1996zw,Heinrich:2010ax}). For processes with up to four external particles, this usually happens near the edge of phase space \cite{Denner:2005nn}, which is consistent with our findings.}.

In the conventional VMD model,
pseudoscalar mesons do not couple directly to photons
but through the exchange of intermediate vectors.
Thus, in this framework, a particular $VP\gamma$ coupling constant
times its normalised form factor, cf.~Eq.~(\ref{eqeffvertex}),
is given by\footnote{Should $q^2$ be timelike, that is, $q^2>0$, then
an imaginary part would need to be added to the propagator; this introduces the associated resonance width effects and rids the propagator from
its divergent behaviour.}
\begin{equation}
\label{VMDcouplings}
g_{VP\gamma}\,\hat F_{VP\gamma}(q^2)=
\sum_{V^\prime}\frac{g_{VV^\prime P}\,g_{V^\prime\gamma}}{M_{V^\prime}^2-q^2}\ ,
\end{equation}
where $g_{VV^\prime P}$ are the vector-vector-pseudoscalar couplings,
$g_{V^\prime\gamma}$ the vector-photon conversion couplings,
and $M_{V^\prime}$ the intermediate vector masses.
In the $SU(3)$-flavour symmetry and OZI-rule respecting limits, 
one could express all the $g_{VP\gamma}$ in terms of
a single coupling constant and $SU(3)$-group factors \cite{Bramon:1994pq}.
However, to 
account for the unavoidable $SU(3)$-flavour symmetry-breaking and
OZI-rule violating effects, we make use of the simple, yet powerful,
phenomenological quark-based model first presented in Ref.~\cite{Bramon:2000fr},
which was developed to describe $V\to P\gamma$ and $P\to V\gamma$ radiative decays.
According to this model, the decay couplings can be expressed as
\begin{equation}
\begin{aligned}
g_{\rho^0\pi^0\gamma} &= \frac{1}{3}g\ ,\\
g_{\rho^0\eta\gamma} &= gz_{\textrm{NS}}\cos{\phi_P}\ ,\\
g_{\rho^0\eta^{\prime}\gamma} &= gz_{\textrm{NS}}\sin{\phi_P}\ ,\\
g_{\omega \pi^0 \gamma} &= g\cos{\phi_V}\ ,\\
g_{\omega\eta\gamma} &= \frac{1}{3}g\Big(z_{\textrm{NS}}\cos{\phi_P}\cos{\phi_V} 
- 2 \frac{\overline{m}}{m_s}z_{\textrm{S}}\sin{\phi_P}\sin{\phi_V}\Big)\ ,\\
g_{\omega\eta^{\prime}\gamma} &= \frac{1}{3}g\Big(z_{\textrm{NS}}\sin{\phi_P}\cos{\phi_V}
+ 2 \frac{\overline{m}}{m_s}z_{\textrm{S}}\cos{\phi_P}\sin{\phi_V}\Big)\ ,\\
g_{\phi\pi^0\gamma} &= g\sin{\phi_V}\ ,\\
g_{\phi\eta\gamma} &= \frac{1}{3}g\Big(z_{\textrm{NS}}\cos{\phi_P}\sin{\phi_V}
+ 2 \frac{\overline{m}}{m_s}z_{\textrm{S}}\sin{\phi_P}\cos{\phi_V}\Big)\ ,\\
g_{\phi\eta^{\prime}\gamma} &= \frac{1}{3}g\Big(z_{\textrm{NS}}\sin{\phi_P}\sin{\phi_V}
- 2 \frac{\overline{m}}{m_s}z_{\textrm{S}}\cos{\phi_P}\cos{\phi_V}\Big)\ ,
\end{aligned}
\label{eqcoups}
\end{equation}
where $g$ is a generic electromagnetic constant,
$\phi_P$ is the pseudoscalar $\eta$-$\eta^{\prime}$ mixing 
angle in the quark-flavour basis, 
$\phi_V$ is the vector $\omega$-$\phi$ mixing angle in the same basis,
$\overline{m}/m_s$ is the quotient of constituent quark masses, and
$z_{\textrm{NS}}$ and $z_{\textrm{S}}$ are the \textit{non-strange} and \textit{strange} 
multiplicative factors accounting for the relative meson wavefunction overlaps 
\cite{Bramon:2000fr,Escribano:2020jdy}.
By performing an optimisation fit to the most 
up-to-date $VP\gamma$ experimental data 
\cite{Zyla:2020zbs}, one can find values for the above parameters
\begin{equation}
\begin{gathered}
\begin{aligned}
g &= 0.70 \pm 0.01 \ \textrm{GeV}^{-1} \ ,	&	
z_{\textrm{S}}\overline{m}/m_s = 0.65 \pm 0.01 \ ,\\
\phi_{P} &= (41.4 \pm 0.5)^\circ \ ,	&	
\phi_{V} = (3.3 \pm 0.1)^\circ \ ,\\
\end{aligned}
\\
z_{\textrm{NS}} = 0.83 \pm 0.02 \ .
\end{gathered}
\label{eqfit}
\end{equation}

Given the very wide decay width of the $\rho^0$ resonance, which, in turn,
is associated to its very short lifetime, the use of the usual Breit-Wigner approximation for
the $\rho^0$ propagator is not justified.
Instead, an energy-dependent width for the vector propagator ought to be considered, which may be written for a generic $\hat{q}^2$ as follows
\begin{equation}
\label{eqEneProp}
\Gamma_{\rho^0}(\hat{q}^2) = 
\Gamma_{\rho^0}\times\Bigg(\frac{\hat{q}^2-4m_{\pi^{\pm}}^2}{m_{\rho^0}^2
-4m_{\pi^{\pm}}^2}\Bigg)^{3/2}\times\theta(\hat{q}^2-4m_{\pi^{\pm}}^2)\ ,
\end{equation}
where $\theta(x)$ is the Heaviside step function. Strictly, one would now need to plug Eq.~(\ref{eqEneProp}) into Eq.~(\ref{eqamp1}) and perform the loop integral, which represents a computation challenge in its own right and is outside of the scope of the present work.\footnote{One could write, for example, the $\rho^0$ energy-dependent propagator $f(s)=\frac{m_{\rho}^2}{m_{\rho}^2-s-i m_{\rho} \Gamma_{\rho}(s)}$ as a once-subtracted dispersion relation, $f(s)=f(s_0)+\frac{s-s_0}{\pi}\int_{s_{\textrm{th}}}^\infty \frac{\textrm{Im}f(s^{\prime})\ \textrm{d}s^{\prime}}{(s^{\prime}-s_0)(s^{\prime}-s-i\epsilon)}$, where $s_{\textrm{th}}$ is the particle production threshold, in the case at hand $s_{\textrm{th}}=4m_{\pi}^2$, and $s_0$ is the subtraction point such that $s_0<s_{\textrm{th}}$, e.g.~$s_0=0$. One would then perform the loop integral in the usual way, leaving the dispersion integral to the end of the computation.} 
With this in mind, and for the sake of simplicity, we resolve to stick with the Breit-Wigner approximation for the $\rho^0$ propagator despite being a potential source of error. The energy-dependent propagator is not needed, though, for the $\omega$ and $\phi$ resonances, as their associated decay widths are narrow and, therefore, use of the usual Breit-Wigner approximation suffices.

Using the most recent empirical data for the meson masses and total decay widths from 
Ref.~\cite{Zyla:2020zbs}, together with all the above considerations,
one arrives at the decay width results shown in Table~\ref{tab1} for the six 
$\eta^{(\prime)}\to\pi^0l^+l^-$ and $\eta^{\prime}\to\eta l^+l^-$ processes.
The total decay widths associated to the electron modes turn out to be larger than
the ones corresponding to the muon modes despite the second and third terms in the
unpolarised squared amplitude (cf.~Eq.~(\ref{eqsqamp}))
being helicity suppressed for the electron modes.
This suppression, though, does not overcome the phase space suppression for the muon modes, 
yielding $\Gamma(\scaleto{\eta^{(\prime)}\to\pi^0}{10pt}e^+e^-)>
\Gamma(\scaleto{\eta^{(\prime)}\to\pi^0\mu^+\mu^-}{10pt})$ and 
$\Gamma(\scaleto{\eta^{\prime}\to\eta}{9pt}e^+e^-)>
\Gamma(\scaleto{\eta^{\prime}\to\eta\mu^+\mu^-}{9pt})$.

\begin{table*}[b]
\centering
{\def\arraystretch{1.2}\tabcolsep=45pt
\begin{tabular}[c]
{@{\hskip 0.05in}c @{\hskip 0.2in}c @{\hskip 0.2in}c @{\hskip 0.2in}c @{\hskip 0.05in}}
\hline\\[-13pt] \hline
Decay & $\Gamma_{\textrm{th}}$ & BR$_{\textrm{th}}$ & BR$_{\textrm{exp}}$\\ 
\hline\\[-13pt] \hline
$\eta\to\pi^0e^+e^-$ & $2.7(1)(1)(2)\times 10^{-6}$ eV 
& $2.0(1)(1)(1)\times 10^{-9}$ & 
$<\ 7.5\times 10^{-6}$ (CL=$90\%$)~\cite{Adlarson:2018imw}\\
$\eta\to\pi^0\mu^+\mu^-$ & $1.4(1)(1)(1)\times 10^{-6}$ eV
& $1.1(1)(1)(1)\times 10^{-9}$ & 
$<\ 5\times 10^{-6}$ (CL=$90\%$)~\cite{Zyla:2020zbs}\\
\hline
$\eta^{\prime}\to\pi^0e^+e^-$ & $8.7(5)(6)(6)\times 10^{-4}$ eV
& $4.5(3)(4)(4)\times 10^{-9}$ & 
$<\ 1.4\times 10^{-3}$ (CL=$90\%$)~\cite{Zyla:2020zbs}\\
$\eta^{\prime}\to\pi^0\mu^+\mu^-$ & $3.3(2)(4)(3)\times 10^{-4}$ eV
& $1.7(1)(2)(2)\times 10^{-9}$ & 
$<\ 6.0\times 10^{-5}$ (CL=$90\%$)~\cite{Zyla:2020zbs}\\
\hline
$\eta^{\prime}\to\eta e^+e^-$ & $8.3(0.5)(0.1)(3.5)\times 10^{-5}$ eV
& $4.3(0.3)(0.2)(1.8)\times 10^{-10}$ & 
$<\ 2.4\times 10^{-3}$ (CL=$90\%$)~\cite{Zyla:2020zbs}\\
$\eta^{\prime}\to\eta\mu^+\mu^-$ & $3.0(0.2)(0.1)(1.1)\times 10^{-5}$ eV
& $1.5(1)(1)(5)\times 10^{-10}$ &
$<\ 1.5\times 10^{-5}$ (CL=$90\%$)~\cite{Zyla:2020zbs}\\
\hline\\[-13pt] \hline
\end{tabular}
}
\caption{Decay widths and branching ratios for the six $C$-conserving decays 
$\eta^{(\prime)}\to\pi^0l^+l^-$ and $\eta^{\prime}\to\eta l^+l^-$ ($l=e$ or $\mu$).
First error is experimental, second is down to numerical integration and
third is due to model dependency.}
\label{tab1}
\end{table*}

Let us now look at the contributions from the different vector meson exchanges to the total decay widths.
For the first decay, i.e.~$\eta\to\pi^0e^+e^-$, we find that the contribution from the $\rho^0$ exchange is $\sim 26\%$, the contribution from the $\omega$ is $\sim 22\%$, whilst the one from the $\phi$ is negligible, i.e.~$\sim 0\%$. The interference between the $\rho^0$ and the $\omega$ is constructive, accounting for the $\sim 49\%$; similarly, the interference between the $\rho^0$ and the $\omega$ with the $\phi$ is constructive and about $\sim 3\%$. 
The contributions to the second decay, i.e.~$\eta\to\pi^0\mu^+\mu^-$, are $\sim 25\%$, $\sim 23\%$ and $\sim 0\%$ from the $\rho^0$, $\omega$, and $\phi$ exchanges, respectively. As before, the interference between the $\rho^0$ and the $\omega$ is constructive, weighing $\sim 48\%$, and the interference between the $\rho^0$ and the $\omega$ with the $\phi$ is constructive and accounts for approximately the $\sim 4\%$.
For the third decay, i.e.~$\eta^{\prime}\to\pi^0e^+e^-$, the contributions from the $\rho^0$, $\omega$ and $\phi$ turn out to be $\sim 16\%$, $\sim 39\%$ and $\sim 0\%$, respectively; the interference between the $\rho^0$ and the $\omega$ exchanges is constructive and accounts for the $\sim 47\%$, whilst the interference between the $\rho^0$ and $\omega$ with the $\phi$ is destructive and weighs approximately $\sim 2\%$. 
The contributions to the fourth decay, i.e.~$\eta^{\prime}\to\pi^0\mu^+\mu^-$, from the $\rho^0$, $\omega$ and $\phi$ exchanges are $\sim 20\%$, $\sim 35\%$ and $\sim 0\%$, respectively. The interference between the $\rho^0$ and the $\omega$ is constructive, representing a $\sim 53\%$ contribution, whilst the interference between the $\rho^0$ and the $\omega$ with the $\phi$ is destructive and accounts for the $\sim 8\%$.
The fifth decay, i.e.~$\eta^{\prime}\to\eta e^+e^-$, gets contributions from the exchange of $\rho^0$, $\omega$ and $\phi$ resonances of approximately $\sim 76\%$, $\sim 1\%$ and $\sim 2\%$, respectively; the interference between the $\rho^0$ and the $\omega$ is constructive weighing $\sim 29\%$, and the interference between the $\rho^0$ and the $\omega$ with the $\phi$ is destructive and contributes with roughly the $\sim 8\%$.
Finally, for the sixth decay, i.e.~$\eta^{\prime}\to\eta \mu^+\mu^-$, we find that the contribution from the $\rho^0$ exchange is $\sim 94\%$, the contribution from the $\omega$ is $\sim 2\%$ and the one from the $\phi$ is $\sim 3\%$; the interference between the $\rho^0$ and the $\omega$ is constructive and accounts for the $\sim 26\%$, whilst the interference between the $\rho^0$ and the $\omega$ with the $\phi$ is destructive weighing close to $\sim 25\%$.
The tiny contribution from the $\phi$ exchange to the decay widths of the six processes is explained by the relatively small product of VMD $VP\gamma$ coupling constants. Likewise, the comparatively minute contribution from the $\omega$ exchange to the decay widths of the last two reactions is down to the significantly smaller product of coupling constants, if compared to that of the $\rho^0$ exchange.

In order to assess a systematic error associated to the model dependency of our predictions, we repeat all the above calculations in the context of
Resonance Chiral Theory (RChT).
In this framework, the $VP\gamma$ effective vertex is made of two contributions, a local $VP\gamma$ vertex weighted by a coupling constant, $h_V$, and a non-local one built from the exchange of an intermediate vector which, again, is weighted by a second coupling constant, $\sigma_V$, times the vector-photon conversion factor $f_V$.
For a given $VP\gamma$ transition,
this effective vertex can be written in the
$SU(3)$-flavour symmetry limit as \cite{Prades:1993ys} 
\begin{equation}
\label{RChTcouplings}
g_{VP\gamma}\,\hat F_{VP\gamma}(q^2)=
C_{VP\gamma}\ \lvert e\rvert \frac{4\sqrt{2}\,h_V}{f_\pi} 
\left(1+\frac{\sigma_V f_V}{\sqrt{2}\,h_V}
\frac{q^2}{M_{V^\prime}^2-q^2}\right)\ ,
\end{equation}
where $C_{VP\gamma}$ are $SU(3)$-group factors
and, depending on the process, the exchanged vector is or is not the same as the initial vector (see Refs.~\cite{Prades:1993ys,Eidelman:2010ta}
for each particular case).
To fix the $VP\gamma$ couplings in this second approach,
we make use of the extended Nambu--Jona-Lasinio (ENJL) model,
where $h_V$ is found to be $h_V=0.035$ \cite{Prades:1993ys}.
The $VVP$ coupling $\sigma_V$ obtained using the ENJL model 
turns out to be $\sigma_V=0.28$.
However, $\sigma_V$ can also be obtained from the analysis of the
dilepton mass spectrum in $\omega\to\pi^0\mu^+\mu^-$ decays,
where one finds $\sigma_V\approx 0.58$ \cite{Ivashyn:2011hb}.
Due to the fact that $\sigma_V$ is poorly known and the 
dispersion of the above estimations is large,
we do not consider the $q^2$ dependence of the form factors
in the subsequent calculations.
An alternative model to fix $g_{VP\gamma}$, the normalisation of the form factors,
is the Hidden Gauge Symmetry (HGS) model \cite{Bando:1987br},
where the vector mesons are considered as gauge bosons of a hidden symmetry.
Within this model, a $VP\gamma$ transition proceeds uniquely through the 
exchange of intermediate vector mesons.
In this sense, it is equivalent to the conventional VMD model
with the relevant exception of including direct $\gamma P^3$ terms
($P$ being a pseudoscalar meson), which are forbidden in VMD
\cite{Fujiwara:1984mp}.
Due to this similarity, we will not make use of 
the HGS model to assess the systematic model error 
and refer the interested reader to 
Ref.~\cite{Bramon:1994pq} for a detailed calculation of the
$g_{VP\gamma}$ couplings in this model.

Next, our results for the semileptonic decays
$\eta^{(\prime)}\to\pi^0l^+l^-$ and $\eta^\prime\to\eta l^+l^-$
in the conventional VMD framework using the $VP\gamma$ couplings
from the phenomenological quark-based model in Eq.~(\ref{eqcoups})
are discussed and, if available,
compared with previous literature.
These predictions include a first experimental error ascribed to the
propagation of the parametric errors in Eq.~(\ref{eqfit}),
a second error down to the numerical integration,
and a third systematic error associated to the model dependence of our approach.
The latter is calculated as the absolute difference between the
predicted central values obtained from the VMD and RChT frameworks
(cf.~Table~\ref{tab1}).

\sloppy
Our prediction for the decay width 
$\Gamma(\scaleto{\eta\to\pi^0}{10pt}e^+e^-)=(2.7\pm 0.1\pm 0.1\pm 0.2)\times 10^{-6}$ eV
is about an order of magnitude smaller than the one provided by Cheng in 
Ref.~\cite{Cheng:1967zza} (cf.~section~\ref{Intro}), 
i.e.~$\Gamma(\scaleto{\eta\to\pi^0}{10pt}e^+e^-)=1.3\times 10^{-5}$ eV;
however, by plugging into our expressions the couplings that Cheng used in his work,
we find a decay width $\Gamma(\scaleto{\eta\to\pi^0}{10pt}e^+e^-)\approx 2.8\times10^{-5}$ eV, 
which is about a factor of two larger than Cheng's result,
and the difference might be down to the simplifications that he had to carry out
in his calculations, as well as the propagators that have been employed in
the present work\footnote{Note that in Ref.~\cite{Cheng:1967zza}
Cheng used vector propagators without total decay widths (i.e.~Feynman propagators)
for the vector exchanges whilst we use Breit-Wigner propagators.}. 
In addition,
from our calculations one can also get a prediction for the ratio of
branching ratios\footnote{Here,
we are using the experimental value for the decay width 
$\Gamma(\scaleto{\eta\to\pi^0\gamma\gamma}{10pt})$
provided in Ref.~\cite{Zyla:2020zbs}.
Alternatively, one could have used the theoretical prediction from 
Ref.~\cite{Escribano:2018cwg}
$\Gamma(\scaleto{\eta\to\pi^0\gamma\gamma}{10pt})
\vert_{\scaleto{\textrm{th}}{4pt}}=0.17\pm 0.01$ eV to obtain 
$\Gamma(\scaleto{\eta\to\pi^0}{10pt}e^+e^-)/\Gamma(\scaleto{\eta\to\pi^0\gamma\gamma}{10pt})
=(1.6\pm 0.3)\times 10^{-5}$.} 
$\Gamma(\scaleto{\eta\to\pi^0}{10pt}e^+e^-)/\Gamma(\scaleto{\eta\to\pi^0\gamma\gamma}{10pt})
=(8.0\pm 1.6)\times 10^{-6}$,
which is not far from Cheng's model independent estimation of 
$\Gamma(\scaleto{\eta\to\pi^0}{10pt}e^+e^-)/\Gamma(\scaleto{\eta\to\pi^0\gamma\gamma}{10pt})
\approx 10^{-5}$,
but more than two orders of magnitude larger than Smith's result\footnote{The
discrepancy with Smith's relative branching ratio 
might be explained, though, by the effect of
$p$-wave terms that he neglected after admitting that they are not necessarily small.}
$\Gamma(\scaleto{\eta\to\pi^0}{10pt}e^+e^-)/\Gamma(\scaleto{\eta\to\pi^0\gamma\gamma}{10pt})
=3.6\times 10^{-8}$ \cite{Smith:1968erc};
as well as this, for the muon mode we find the 
relative branching ratio\footnote{Once again, 
should we have used the theoretical prediction from Ref.~\cite{Escribano:2018cwg} 
$\Gamma(\scaleto{\eta\to\pi^0\gamma\gamma}{10pt})\vert_{\scaleto{\textrm{th}}{4pt}}
=0.17\pm 0.01$ eV, we would have obtained 
$\Gamma(\scaleto{\eta\to\pi^0}{10pt}\mu^+\mu^-)/\Gamma(\scaleto{\eta\to\pi^0\gamma\gamma}{10pt})
=(8.3\pm 1.5)\times 10^{-6}$.} 
$\Gamma(\scaleto{\eta\to\pi^0\mu^+\mu^-}{10pt})/\Gamma(\scaleto{\eta\to\pi^0\gamma\gamma}{10pt})
=(4.2\pm 0.8)\times 10^{-6}$,
which is roughly an order of magnitude smaller than Smith's estimation 
$\Gamma(\scaleto{\eta\to\pi^0\mu^+\mu^-}{10pt})/\Gamma(\scaleto{\eta\to\pi^0\gamma\gamma}{10pt})
=6.0\times 10^{-5}$ \cite{Smith:1968erc}. 
Moreover,
our decay widths for both the $\eta\to\pi^0e^+e^-$ and $\eta\to\pi^0\mu^+\mu^-$ processes
are consistent with the lower bounds provided by Ng et al.~in Ref.~\cite{Ng:1992yg}, 
i.e.~$\Gamma(\scaleto{\eta\to\pi^0}{10pt}e^+e^-)\vert_{\scaleto{\textrm{VMD}}{4pt}}
=1.1_{\scaleto{-0.5}{5pt}}^{\scaleto{+0.6}{5pt}}\ \mu$eV and 
$\Gamma(\scaleto{\eta\to\pi^0\mu^+\mu^-}{10pt})\vert_{\scaleto{\textrm{VMD}}{4pt}}
=0.5_{\scaleto{-0.2}{5pt}}^{\scaleto{+0.3}{5pt}}\ \mu$eV making use of VMD, and 
$\Gamma(\scaleto{\eta\to\pi^0\mu^+\mu^-}{10pt})\vert_{\scaleto{\textrm{constr}}{4pt}}
=0.9_{\scaleto{-0.5}{5pt}}^{\scaleto{+0.6}{5pt}}\ \mu$eV and 
$\Gamma(\scaleto{\eta\to\pi^0\mu^+\mu^-}{10pt})\vert_{\scaleto{\textrm{destr}}{5pt}}
=0.3_{\scaleto{-0.2}{5pt}}^{\scaleto{+0.4}{5pt}}\ \mu$eV
using VMD supplemented with the exchange of an $a_0$ scalar meson.
Using the quark-box diagram and a constituent quark mass $m=330$ MeV$/c^2$,
Ng et al.~provided in Ref.~\cite{Ng:1993sc} an estimation for the electron mode, 
$\Gamma(\scaleto{\eta\to\pi^0}{10pt}e^+e^-)\vert_{\scaleto{\textrm{box}}{4pt}}
\geq1.2\pm0.2\ \mu$eV,
which is in accordance with our result, and an estimate for the muon mode, 
$\Gamma(\scaleto{\eta\to\pi^0\mu^+\mu^-}{10pt})\vert_{\scaleto{\textrm{box}}{4pt}}
\geq4.3\pm0.7\ \mu$eV,
which in this case is incompatible with our calculation\footnote{Note,
however, that, as part of their calculation, they had to 
estimate the decay width of the $\eta\to\pi^0\gamma\gamma$ process 
using their quark-box model and found 
$\Gamma(\scaleto{\eta\to\pi^0\gamma\gamma}{10pt}) = 0.60\pm0.10$ eV
for a constituent quark mass $m=330$ MeV$/c^2$,
which is approximately a factor of two larger than the 
current experimental measurement. 
Therefore, it is no surprise that their estimates for the associated
$\eta\to\pi^0 l^+l^-$ processes are at the upper end of the spectrum of estimations.}.
Additionally, Ng et al.~also presented in Ref.~\cite{Ng:1993sc}
a recalculation of their previous VMD results from Ref.~\cite{Ng:1992yg}, yielding 
$\Gamma(\scaleto{\eta\to\pi^0}{10pt}e^+e^-)\vert_{\scaleto{\textrm{VMD}}{4pt}}
\geq3.5\pm0.8\ \mu$eV and 
$\Gamma(\scaleto{\eta\to\pi^0\mu^+\mu^-}{10pt})\vert_{\scaleto{\textrm{VMD}}{4pt}}
\geq2.4\pm0.8\ \mu$eV,
which are consistent with our results if one considers the associated errors. 
Our decay width calculations for the other four processes,
i.e.~$\eta^{\prime}\to\pi^0l^+l^-$ and $\eta^{\prime}\to\eta l^+l^-$,
cannot be compared with any previously published theoretical results,
as these decays have been calculated, to the best of our knowledge,
for the first time in the present work. 
Likewise, comparison with the most up-to-date empirical data provides limited value
given that the corresponding current experimental upper bounds,
though consistent with our theoretical predictions,
are many orders of magnitude larger (cf.~Table~\ref{tab1}).

\begin{figure*}[t!]
\centering
\begin{subfigure}[b]{0.49\textwidth}
	  \centering
        \includegraphics[height=0.23\textheight,width=0.9\textwidth,
        trim=7mm 0mm 0mm 0mm, clip=false]{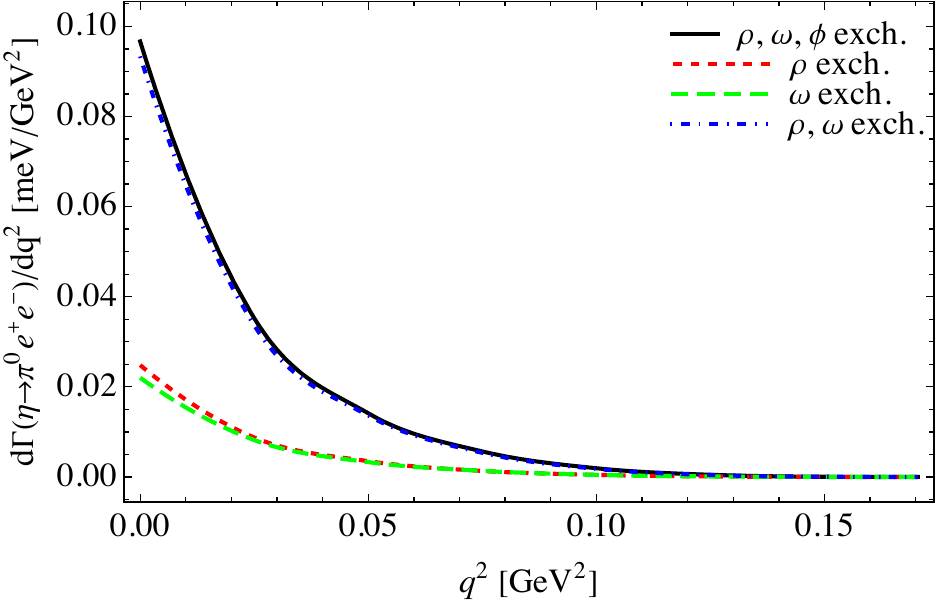}
        \caption{$\eta\to\pi^0e^+e^-$}
        \label{fig2_a}
\end{subfigure}
\begin{subfigure}[b]{0.49\textwidth}
	  \centering
        \includegraphics[height=0.23\textheight,width=0.9\textwidth,
        trim=0mm 0mm 0mm 0mm, clip=false]{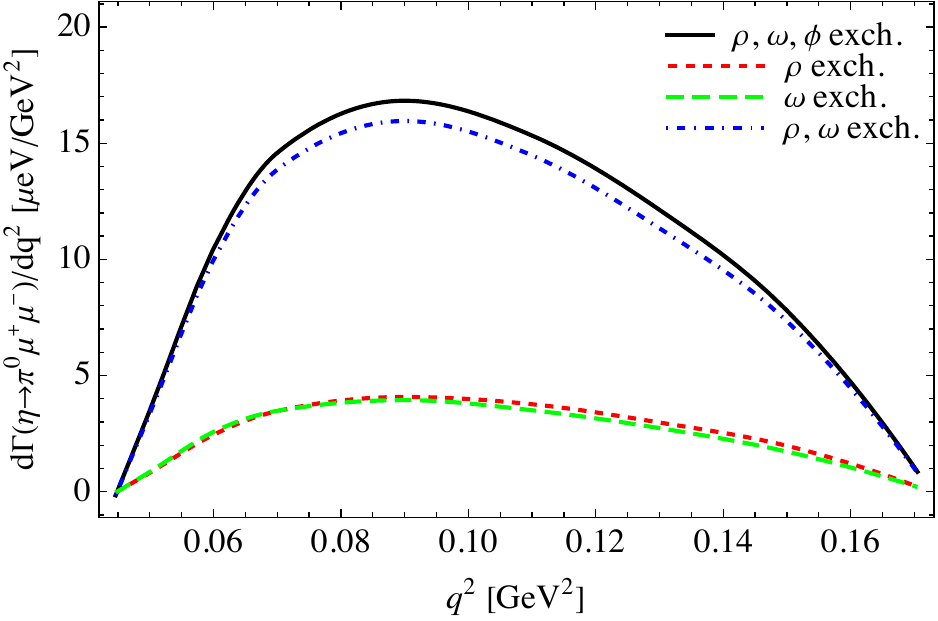}
        \caption{$\eta\to\pi^0\mu^+\mu^-$}
        \label{fig2_b}
\end{subfigure} \\[3\tabcolsep]
\begin{subfigure}[t]{0.49\textwidth}
	  \centering
        \includegraphics[height=0.23\textheight,width=0.9\textwidth,
        trim=1.5mm 0mm 0mm 0mm, clip=false]{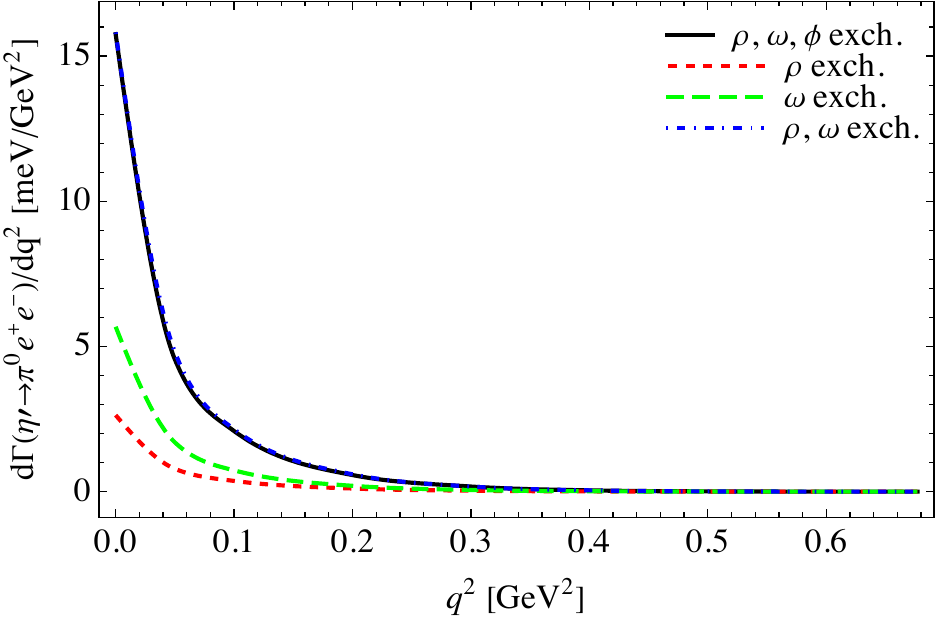}
        \caption{$\eta^{\prime}\to\pi^0e^+e^-$}
        \label{fig2_c}
\end{subfigure} 
\begin{subfigure}[t]{0.49\textwidth}
	  \centering
        \includegraphics[height=0.23\textheight,width=0.9\textwidth,
        trim=2mm 0mm 0mm 0mm, clip=false]{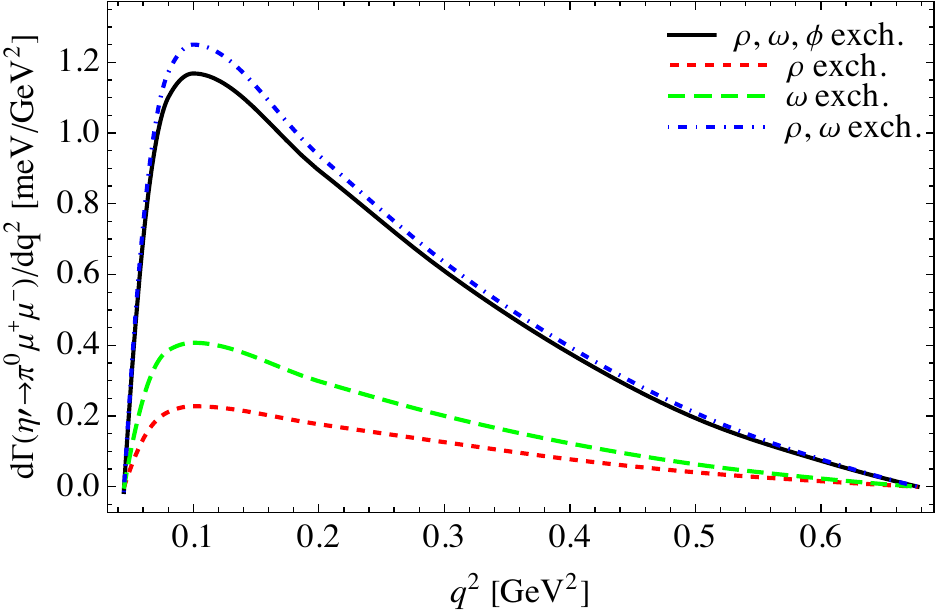}
        \caption{$\eta^{\prime}\to\pi^0\mu^+\mu^-$}
        \label{fig2_d}
\end{subfigure} \\[3\tabcolsep]
\begin{subfigure}[t]{0.49\textwidth}
	  \centering
        \includegraphics[height=0.23\textheight,width=0.9\textwidth,
        trim=-2.5mm 0mm 0mm 0mm, clip=false]{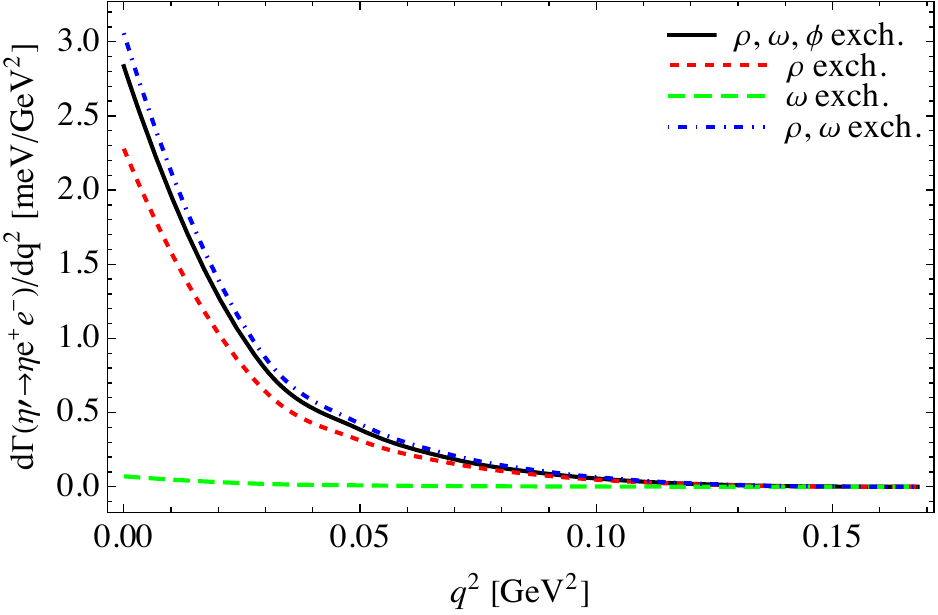}
        \caption{$\eta^{\prime}\to\eta e^+e^-$}
        \label{fig2_e}
\end{subfigure}
\begin{subfigure}[t]{0.49\textwidth}
	  \centering
        \includegraphics[height=0.23\textheight,width=0.9\textwidth,
        trim=2mm 0mm 0mm 0mm, clip=false]{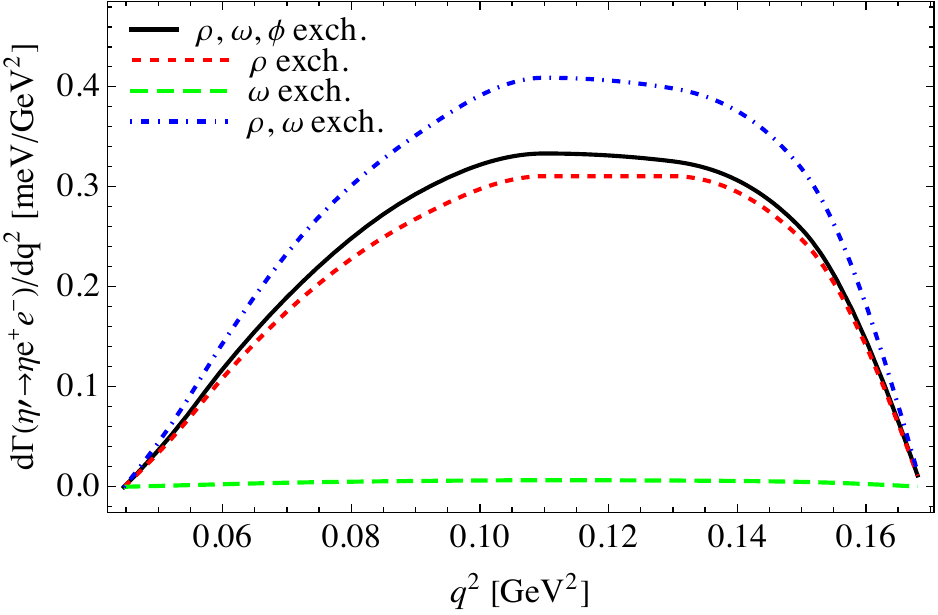}
        \caption{$\eta^{\prime}\to\eta\mu^+\mu^-$}
        \label{fig2_f}
\end{subfigure} \\[2\tabcolsep]
\caption{
Dilepton energy spectra corresponding to the six $C$-conserving semileptonic decay processes
$\eta^{(\prime)}\to\pi^0 l^+l^-$ and $\eta^{\prime}\to\eta l^+l^-$ ($l=e$ or $\mu$)
as a function of the dilepton invariant mass $q^2$.}\label{fig2}
\end{figure*}

Finally, theoretical results for the dilepton energy spectra of the six
$C$-conserving semileptonic decays 
are presented in Fig.~\ref{fig2}.
The energy spectra for the three electron modes,
which are displayed in Fig.~\ref{fig2} (a), (c) and (e),
take off as the dilepton invariant mass $q^2$ approaches zero.
This is in line with Cheng's and Ng et al.'s energy spectra for the
$\eta\to\pi^0 e^+e^-$ (Refs.~\cite{Cheng:1967zza} and \cite{Ng:1992yg}, respectively),
which exhibit the same behaviour at low $q^2$.
It appears as though the electron modes \textit{prefer}
to proceed through the emission of (relativistic) collinear electron-positron pairs 
(i.e.~$\theta_{e^+e^-}\simeq 0$, where $\theta_{e^+e^-}$ is the electron-positron angle).
The reason for this can easily be understood from dynamics\footnote{It must be noted,
though, that the kinematics of the electron modes also contribute to
this particular shape of the energy spectra, producing a somewhat synergistic effect.}
if one assumes the electron and positron to be massless, $m_e\approx 0$;
then, by inspection of Eq.~(\ref{eqsqamp}),
one can determine that the unpolarised squared amplitude is maximised when $q^2\to 0$,
which occurs when $\theta_{e^+e^-}\simeq 0$\footnote{Note that
$q^2\simeq 2p_{e^+}p_{e^-}\simeq 
2|\boldsymbol{p_{e^+}}||\boldsymbol{p_{e^-}}|(1-\cos{\theta_{e^+e^-}})$
in the leptonic massless limit, i.e.~$m_e\approx 0$.}.
Physically, it may be explained to some extent by the fact that the
diphoton invariant spectra for the three
$\eta^{(\prime)}\to\pi^0\gamma\gamma$ and $\eta^{\prime}\to\eta\gamma\gamma$
peak at low $m_{\gamma\gamma}^2$
(cf., e.g., Ref.~\cite{Escribano:2018cwg} and references therein).
On the other hand, the dilepton energy spectra for the muon modes,
shown in Fig.~\ref{fig2} (b), (d) and (f), are bell-shaped,
which is driven by the kinematics of the processes.
This, once more, seems to be consistent with Ng et al.'s \cite{Ng:1992yg}
energy spectrum for the $\eta\to\pi^0 \mu^+\mu^-$.
It is interesting to observe that the energy spectra for the
$\eta\to\pi^0\mu^+\mu^-$ and $\eta^{\prime}\to\pi^0 \mu^+\mu^-$
are skewed to the left (i.e.~small values for $\theta_{\mu^+\mu^-}$
are favoured, where $\theta_{\mu^+\mu^-}$ is the muon-antimuon angle),
whilst the energy spectrum for the $\eta^{\prime}\to\eta \mu^+\mu^-$
is skewed to the right (i.e.~somewhat larger values for $\theta_{\mu^+\mu^-}$ are preferred). 
This is more difficult to explain\footnote{A
qualitative explanation could be given from a statistical mechanics viewpoint,
whereby high momentum $\eta$ mesons in the final state would be Boltzmann suppressed
compared to high momentum $\pi^0$ states.}
given that this effect, which is connected to the fact that $m_{\eta}>m_{\pi^0}$,
is a consequence of the complex dynamical interplay between the different terms in 
Eq.~(\ref{eqsqamp}).
Surprisingly, the kinematics of the reactions do not seem to play a significant role
in this difference in skewness.


\section{\label{Conclusions}Conclusions}

In this work, the $C$-conserving decay modes
$\eta^{(\prime)}\to\pi^0l^+l^-$ and $\eta^{\prime}\to\eta l^+l^-$ ($l=e$ or $\mu$)
have been analysed within the theoretical framework of the VMD model.
The associated decay widths and dilepton energy spectra 
have been calculated and presented for the six decay processes.
To the best of our knowledge, the theoretical predictions for the four 
$\eta^{\prime}\to\pi^0l^+l^-$ and $\eta^{\prime}\to\eta l^+l^-$ reactions
that we have provided in this work are the first predictions from theory
that have been published.

The decay width results that we have obtained from our calculations,
which are summarised in Table~\ref{tab1},
have been compared with those available in the published literature.
In general, the agreement is reasonably good considering that the previous analysis
either contain important approximations or consist of unitary lower bounds.
Predictions for the dilepton energy spectra have also been presented
for all the above processes, cf.~Fig.~\ref{fig2}.

Experimental measurements to date have provided upper limits to the decay processes studied in this work.
These upper limits, though,
are still many orders of magnitude larger than the theoretical results that we have presented. 
For this reason, we would like to encourage experimental groups,
such as the WASA-at-COSY and REDTOP Collaborations,
to study these semileptonic decays processes,
as we believe that they can represent a fruitful arena in the search for
new physics beyond the Standard Model.


\begin{acknowledgements}

We would like to thank Sergi Gonz\`{a}lez-Sol\'{i}s
for suggesting us to work on this particular topic and 
Pablo Roig for recommending relevant literature on Resonance Chiral Theory.
As well as this, we are very grateful to Pablo Sanchez-Puertas
for a thorough read of the manuscript and for checking some of the numerical results.
Finally, this work is supported by the
Secretaria d'Universitats i Recerca del Departament d'Empresa i Coneixement
de la Generalitat de Catalunya under the grant 2017SGR1069,
by the Ministerio de Econom\'{i}a, Industria y Competitividad under the grant FPA2017-86989-P, 
from the Centro de Excelencia Severo Ochoa under the grant 
SEV-2016-0588 and from the EU
STRONG-2020 project under the program H2020-INFRAIA-2018-1, 
grant agreement No. 824093. IFAE is partially funded by the CERCA 
program of the Generalitat de Catalunya.

\end{acknowledgements}



\appendix

\section{\label{appA} Definition of parameters $\pmb{A_i}$, $\pmb{B_i}$, $\pmb{C_i}$ and $\pmb{D_i}$}

The parameters $A_i$, $B_i$, $C_i$ and $D_i$ ($i=1,2$) from Eq.~(\ref{eqnums12}) are defined as follows
\begin{fleqn}
\begin{equation}
\label{eqA1}
\begin{aligned}
\quad A_1 & = -(x+y+2z)(P\cdot p_{+})-(y+2z-1)(P\cdot p_{-})
+(-x+y+1)(p_{-}\cdot p_{+}) \\
& \quad +m_{l}^2(2x+y+1)+\frac{3 z m_{\eta^{(\prime)}}^2}{2}\ ,
\end{aligned}
\end{equation}
\end{fleqn}
\begin{fleqn}
\begin{equation}
\label{eqA2}
\begin{aligned}
\quad A_2 & \equiv  -A_1\ \ \mathrm{with}\ \ p_{+}\leftrightarrow p_{-}\ ,
\end{aligned}
\end{equation}
\end{fleqn}
\begin{fleqn}
\begin{equation}
\label{eqB1}
\begin{aligned}
\quad B_1 & = -2z(x+y)(x+y+z-1)(P\cdot p_{+})^2-2y z(y+z-1)
\times (P\cdot p_{-})^2-2x y(x+y-1)(p_{-}\cdot p_{+})^2 \\
& \quad +(P\cdot p_{-})\Big\{-2z\big[x(2y+z-1)+2y(y+z-1)\big](P\cdot p_{+})
+2y(p_{-}\cdot p_{+})(-2x z+x+y+z-1) \\
& \quad +m_l^2\big[x^2(1-2z)+2x y+2y(y+z-1)\big]\Big\}+ (P\cdot p_{+})
\Big\{m_l^2\big[x^2(2z-1)+2x y(2z-1)-2y(y+z-1)\big] \\
& \quad -2y(p_{-}\cdot p_{+})(x+y+z-1)\Big\}-x^3 m_l^2(p_{-}\cdot p_{+}) 
+z m_{\eta^{(\prime)}}^2 \Big\{(p_{-}\cdot p_{+})\big[x(4y+z-1)+4y(y+z-1)\big] \\
& \quad +m_l^2\big[2x^2+x(4y+3z-3)+4y(y+z-1)\big]\Big\} 
+x m_l^4\big[x^2+2x y+2(y-1)y\big], 
\end{aligned}
\end{equation}
\end{fleqn}
\begin{fleqn}
\begin{equation}
\label{eqB2}
\begin{aligned}
\quad B_2 & \equiv -B_1\ \ \mathrm{with}\ \ p_{+}\leftrightarrow p_{-}\ , 
\end{aligned}
\end{equation}
\end{fleqn}
\begin{fleqn}
\begin{equation}
\label{eqC1}
\begin{aligned}
\quad C_1 & = (x+1)(P\cdot p_{-})+(4x+1)(P\cdot p_{+})
-\Big(\frac{5}{2} x+1\Big) m_{\eta^{(\prime)}}^2\ , 
\end{aligned}
\end{equation}
\end{fleqn}
\begin{fleqn}
\begin{equation}
\label{eqC2}
\begin{aligned}
\quad C_2 & \equiv C_1\ \ \mathrm{with}\ \ p_{+}\leftrightarrow p_{-}\ ,
\end{aligned}
\end{equation}
\end{fleqn}
\begin{fleqn}
\begin{equation}
\label{eqD1}
\begin{aligned}
\quad D_1 & = 2\big[x^3+x^2(2y+2z-1)+x y(y+2z)+y(y+z-1)\big] 
\times (P\cdot p_{+})^2+2y(x y+y+z-1)(P\cdot p_{-})^2 \\
& \quad +x(P\cdot p_{+})\Big\{-2(x^2-y^2+y)(p_{-}\cdot p_{+})+m_l^2\big[x^2+2(y-1)y\big]+(z-1)z m_{\eta^{(\prime)}}^2\Big\}+(P\cdot p_{-}) \\
& \quad \times\Big\{x\big[-2(y-1)y
(p_{-}\cdot p_{+}+m_l^2)+x^2m_l^2-(z-1)z m_{\eta^{(\prime)}}^2\big]+4(x+1)y(x+y+z-1)(P\cdot p_{+})\Big\} \\
& \quad -m_{\eta^{(\prime)}}^2\Big\{(p_{-}\cdot p_{+}+m_l^2) \big[x^2(4y+2z-1)+4 x y(y+z)+4y(y+z-1)\big]+2x^3m_l^2\Big\}\ ,
\end{aligned}
\end{equation}
\end{fleqn}
\begin{fleqn}
\begin{equation}
\label{eqD2}
\begin{aligned}
\quad D_2 & \equiv D_1\ \ \mathrm{with}\ \ p_{+}\leftrightarrow p_{-}\ .
\end{aligned}
\end{equation}
\end{fleqn}
%



\begin{thebibliography}{99}
    
\bibitem{Gan:2020aco} 
L.~Gan, B.~Kubis, E.~Passemar and S.~Tulin, 
[arXiv:2007.00664 [hep-ph]]. 

\bibitem{Fang:2017qgz} 
  S.~s.~Fang, A.~Kupsc and D.~h.~Wei,
  Chin.\ Phys.\ C {\bf 42}, no. 4, 042002 (2018)
  [arXiv:1710.05173 [hep-ex]].

\bibitem{Bernstein:1965hj}
  J.~Bernstein, G.~Feinberg and T.~D.~Lee,
  Phys.\ Rev.\  {\bf 139} (1965) B1650.
  
\bibitem{Lee:1965zz}
  T.~D.~Lee,
  Phys.\ Rev.\  {\bf 140} (1965) B959.

\bibitem{Cheng:1967zza}
  T.~P.~Cheng,
  Phys.\ Rev.\  {\bf 162} (1967) 1734.

\bibitem{Smith:1968erc}
  J.~Smith,
  Phys.\ Rev.\  {\bf 166} (1968) 1629.
  %
   
  \bibitem{Cutkosky:1960sp}
  R.~Cutkosky,
  J. Math. Phys. \textbf{1} (1960), 429-433

\bibitem{Ng:1992yg}
  J.~N.~Ng and D.~J.~Peters,
  Phys.\ Rev.\ D {\bf 46} (1992) 5034.

\bibitem{Ng:1993sc}
  J.~N.~Ng and D.~J.~Peters,
  Phys.\ Rev.\ D {\bf 47} (1993) 4939.
  
\bibitem{Adlarson:2018imw}
  P.~Adlarson {\it et al.} [WASA-at-COSY Collaboration],
  Phys.\ Lett.\ B {\bf 784} (2018) 378
  [arXiv:1802.08642 [hep-ex]].
  
\bibitem{Gatto:2019dhj}
  C.~Gatto [REDTOP],
  [arXiv:1910.08505 [physics.ins-det]].
  
\bibitem{Escribano:2018cwg}
R.~Escribano, S.~Gonz\`alez-Sol\'\i{}s, R.~Jora and E.~Royo,
Phys. Rev. D \textbf{102} (2020) no.3, 034026
[arXiv:1812.08454 [hep-ph]].

\bibitem{Prades:1993ys}
J.~Prades,
Z. Phys. C \textbf{63} (1994), 491-506
[erratum: Z. Phys. C \textbf{11} (1999), 571]
[arXiv:hep-ph/9302246 [hep-ph]].
  
\bibitem{Shtabovenko:2016sxi}
  V.~Shtabovenko, R.~Mertig and F.~Orellana,
  Comput.\ Phys.\ Commun.\  {\bf 207} (2016) 432
  [arXiv:1601.01167 [hep-ph]].

\bibitem{Mertig:1990an}
  R.~Mertig, M.~Bohm and A.~Denner,
  Comput.\ Phys.\ Commun.\  {\bf 64} (1991) 345.

\bibitem{Peskin:1995ev}
  M.~E.~Peskin and D.~V.~Schroeder,
  Westview Press (1995).

\bibitem{Schwartz:2013pla}
  M.~D.~Schwartz,
  Cambridge University Press (2014).

\bibitem{Zyla:2020zbs}
P.~A.~Zyla \textit{et al.} [Particle Data Group],
PTEP \textbf{2020} (2020) no.8, 083C01

\bibitem{Press:2007zz}
  W.~H.~Press, S.~A.~Teukolsky, W.~T.~Vetterling and B.~P.~Flannery,
  Cambridge University Press (2007).

\bibitem{Hahn:1998yk}
T.~Hahn and M.~Perez-Victoria,
Comput. Phys. Commun. \textbf{118} (1999), 153-165
[arXiv:hep-ph/9807565 [hep-ph]].


\bibitem{Denner:2005nn}
A.~Denner and S.~Dittmaier,
Nucl. Phys. B \textbf{734} (2006), 62-115
[arXiv:hep-ph/0509141 [hep-ph]].

\bibitem{Campbell:1996zw}
J.~M.~Campbell, E.~W.~N.~Glover and D.~J.~Miller,
Nucl. Phys. B \textbf{498} (1997), 397-442
[arXiv:hep-ph/9612413 [hep-ph]].

\bibitem{Heinrich:2010ax}
G.~Heinrich, G.~Ossola, T.~Reiter and F.~Tramontano,
JHEP \textbf{10} (2010), 105
[arXiv:1008.2441 [hep-ph]].

\bibitem{Bramon:1994pq}
A.~Bramon, A.~Grau and G.~Pancheri,
Phys. Lett. B \textbf{344} (1995), 240-244

\bibitem{Bramon:2000fr}
A.~Bramon, R.~Escribano and M.~Scadron,
Phys.\ Lett.\ B \textbf{503} (2001), 271-276
[arXiv:hep-ph/0012049 [hep-ph]].
  
\bibitem{Escribano:2020jdy}
R.~Escribano and E.~Royo,
Phys. Lett. B \textbf{807} (2020), 135534
[arXiv:2003.08379 [hep-ph]].

\bibitem{Guerrero:1997ku}
F.~Guerrero and A.~Pich,
Phys. Lett. B \textbf{412} (1997), 382-388
[arXiv:hep-ph/9707347 [hep-ph]].

\bibitem{Eidelman:2010ta}
S.~Eidelman, S.~Ivashyn, A.~Korchin, G.~Pancheri and O.~Shekhovtsova,
Eur. Phys. J. C \textbf{69} (2010), 103-118
[arXiv:1003.2141 [hep-ph]].

\bibitem{Ivashyn:2011hb}
S.~Ivashyn,
Prob. Atomic Sci. Technol. \textbf{2012N1} (2012), 179-182
[arXiv:1111.1291 [hep-ph]].

\bibitem{Bando:1987br}
M.~Bando, T.~Kugo and K.~Yamawaki,
Phys. Rept. \textbf{164} (1988), 217-314

\bibitem{Fujiwara:1984mp}
T.~Fujiwara, T.~Kugo, H.~Terao, S.~Uehara and K.~Yamawaki,
Prog. Theor. Phys. \textbf{73} (1985), 926-941

\end{thebibliography}
\end{document}